\documentclass{aa}  

\usepackage{graphicx}
\usepackage{pdflscape}
\usepackage{txfonts}
\usepackage{natbib}
\usepackage{color}
\usepackage{amssymb}
\usepackage{mathrsfs}
\usepackage{multirow}
\usepackage{tablefootnote}

\begin{document}

   \title{A compact gaseous accretion disk in Keplerian rotation around \object{MWC~147}
    \thanks{Based on observations made with ESO Telescopes at the La Silla Paranal Observatory under programme IDs 082.C-0627, 082.C-0893 and 086.C-0684.}}
    
   \author{
    Edward Hone\inst{1},
    Stefan Kraus\inst{1},
    Claire L. Davies\inst{1},
    Alexander Kreplin\inst{1},
    John D. Monnier\inst{2},
    Fabien Baron\inst{3},
    Rafael Millan-Gabet\inst{4},
    Karl-Heinz Hofmann\inst{5},
    Dieter Schertl\inst{5},
    Judit Sturmann\inst{5},
    Laszlo Sturmann\inst{5},
    Theo Ten Brummelaar\inst{6}, and
    Gerd Weigelt\inst{5}
    }
    
   \authorrunning{Hone et al.}
    
   \institute{Department of Physics and Astronomy, University of Exeter, Stocker Road, Exeter, EX4 4QL, UK
   \and
   Department of Astronomy, University of Michigan, Ann Arbor, MI 48109, USA
   \and
   Department of Physics and Astronomy, Georgia State University, Atlanta, GA, USA
   \and
   Infrared Processing and Analysis Center, California Institute of Technology, Pasadena, CA 91125, USA
   \and
   Max-Planck Institut f\"ur Radioastronomie, Auf dem H\"ugel 69, 53121 Bonn, Germany
   \and
   The CHARA Array of Georgia State University, Mount Wilson Observatory, Mount Wilson, CA 91203, USA
   }

   \date{}

 
  \abstract
    {
    The disks around some Herbig Be stars have been observed to be more compact than the expected dust sublimation radius for such objects, with highly refractory dust grains and optically thick gas emission having been proposed as possible explanations for this phenomenon.
    }
    {
    Previously, the "undersized" Herbig Be star \object{MWC~147} was observed with interferometry, with the results indicating the presence of a compact gaseous disk based on the measured wavelength-dependence of near-infrared/mid-infrared visibilities.
    Our aim is to search for direct evidence for the presence of hot gas inside of the expected dust sublimation radius of \object{MWC~147}.
    }
    {
    By combining VLTI/AMBER spectro-interferometry ($R=12\,000$) with CRIRES spectroscopy ($R=100\,000$) we can both spectrally and spatially resolve the Br$\gamma$ line-emitting gas around \object{MWC~147}.
    Additionally, using CHARA/CLIMB enables us to achieve baseline lengths up to 330m, offering ${\sim}2$ times higher angular resolution (and a better position angle coverage) than has previously been achieved with interferometry for \object{MWC~147}.
    To model the continuum we fit our AMBER and CLIMB measurements with a geometric model of an inclined Gaussian distribution as well as a ring model.
    We fit our high resolution spectra and spectro-interferometric data with a kinematic model of a disk in Keplerian rotation.
    }
    {
    Our interferometric visibility modelling of \object{MWC~147} indicates the presence of a compact continuum disk with a close to face-on orientation.
    We model the continuum with an inclined Gaussian, as well as a ring with a radius of 0.60~mas ($0.39$~au) which is well within the expected dust sublimation radius of $1.52$~au.
    We detect no significant change in the measured visibilities across the Br$\gamma$ line, indicating that the line-emitting gas is located in the same region as the continuum-emitting disk.
    Using our differential phase data we construct photocentre displacement vectors across the Br$\gamma$ line, revealing a velocity profile consistent with a rotating disk.
    We fit our AMBER spectro-interferometry data with a kinematic model of a disk in Keplerian rotation, with both the line-emitting and continuum-emitting components of the disk originating from the same compact region close to the central star.
    The presence of line-emitting gas in the same region as the K-band continuum supports the interpretation that the K-band continuum traces an optically-thick gas disk.
    } 
    {
    Our spatially and spectrally resolved observations of \object{MWC~147} reveal that the K-band continuum and Br$\gamma$ emission both originate from a similar region which is $3.9$ times more compact than the expected dust sublimation radius for the star, with Br$\gamma$ emitted from the accretion disk or disk wind region and exhibiting a rotational velocity profile.
    We conclude that we detect the presence of a compact, gaseous accretion disk in Keplerian rotation around \object{MWC~147}
    }

\keywords{stars: formation --
stars: circumstellar matter --
stars: variables: Herbig Ae/Be
ISM: individual objects: \object{MWC~147} --
techniques: interferometric --
techniques: high angular resolution
}

   \maketitle
%

\section{Introduction}
\label{Sec:intro}

Stars are formed from giant molecular clouds in the interstellar medium that collapse under self-gravity.
The collapsing cloud fragments to form individual proto-stellar cores which, as the young stellar object (YSO) evolves and the system's total angular momentum is conserved, further collapse and flatten to form circumstellar accretion disks, commonly referred to as protoplanetary disks \citep{1987Shu}.
As the star formation process takes place the material of the disk is accreted on to the central protostar \citep{2007Bouvier} or is ejected out of the disk plane in bipolar outflows \citep{2007Shang}.
It has been suggested that such outflows play an important role in the process of stellar spin down \citep{2005Matt} by providing an efficient means of angular momentum removal.\\
The processes of mass accretion and ejection take place in the innermost region of the disk, within the dust sublimation radius.
Here the hydrogen gas is heated to extreme temperatures \citep[${\sim}6000\,$K$ - 12\,000\,$K;][]{2001Muzerolle,2004Muzerolle}, where it becomes hot enough to emit in the Br$\gamma$ recombination line.
The origin of the Br$\gamma$ line in Herbig Ae/Be stars is an open question, with both \citet{2006Ferreira} and \citet{2016Tambovtseva} arguing that multiple line-emission mechanisms are required to explain the majority of observations of Br$\gamma$ emission from young stellar objects.\\

The exact mechanisms that drive accretion and mass outflow in the inner regions of protoplanetary disks are the subject of ongoing scientific investigation.
Whilst the magnetospheric accretion paradigm \citep{2007Bouvier} is widely accepted for low-mass T\,Tauri stars, various complex infall geometries have been proposed by different theoretical models to overcome the intense radiation pressure that could halt mass infall for more massive YSOs \citep{2002McKee,2009Krumholz,2010Kuiper}.
Additionally, many different possible scenarios have been proposed to explain the launching of jets and outflows from the inner disk, such as those outlined in \citet{2006Ferreira}.
This work also describes how the characteristic size of the jet-launching region varies for the various different launching scenarios.
For example it is expected that a magneto-centrifugally driven disk wind will be launched from a region that is more extended than the launching region of a stellar wind or X-wind.
\citeauthor{2006Ferreira} concluded that the most likely ejection scenario involves multiple jet-launching components.
This thesis was demonstrated observationally by \citet{2008Krausa} who measured the radial extension of Br$\gamma$ emission for a selection of five Herbig Ae/Be stars, finding that the size of the Br$\gamma$ emission relative to the continuum can differ significantly for different objects in the sample, suggesting that the physical mechanisms traced by the Br$\gamma$ line are varied for different objects.\\

Spectro-interferometry in the Br$\gamma$ line has been employed for several YSOs in order to determine the kinematics of the jet-launching region with high spectral resolution.
Notably, the works of \citet{2011Weigelt}, \citet{2015GarciaLopez} and \citet{2015CarrattioGaratti} all focused on Herbig Be stars, with each independently finding that their data was best fit with a kinematic model of a disk wind.
\citet{2016Kurosawa} also fit their spectro-interferometric data with a disk wind model but noted that the expected emission from the base of the wind strongly resembled a Keplerian rotation velocity field, making it difficult to distinguish between the two line-emitting scenarios.
In their modelling of Br$\gamma$ emission from \object{MWC~297}, \citet{2017Hone} demonstrated that it was possible to distinguish between a disk wind and Keplerian rotation, as it was shown that an out-of-plane velocity component shifts the perceived rotation axis\footnote{The percieved rotation axis is traced by the photocentre displacement at different velocities of the gas-tracing line} away from the disk axis. In a Keplerian rotation case, it would be expected that the axis of rotation is aligned with the axis of the continuum-emitting disk.\\

\object{MWC~147} (\object{HD\,259431}) is a $6.6\,M_\odot$ Herbig Be star (spectral type B6) which is host to a circumstellar accretion disk \citep{1992Hillenbrand,2002Polomski}.
\citet{2004Hernandez} determined that \object{MWC~147} has a luminosity of $10^{3.19} L_{\odot}$ ($1549\,L_{\odot}$) and is at a distance of 800\,pc, however we adopt the new distance of $711{\pm}24$\,pc measured by Gaia \citep{2016Gaia,2018Gaia} using the more robust geometric distance calculation of \citet{2018Bailer-Jones}.
\citet{2003Bouret} used the FUSE UV spectrograph to examine the far-UV line emission originating from \object{MWC~147} and determined the stellar radial velocity to $V_{rad} = 43$\,km/s.
The first in-depth interferometric view of the disk around \object{MWC~147} was provided by \citet{2008Krausb}, using PTI, IOTA, VLTI/MIDI and VLTI/AMBER data.
\citeauthor{2008Krausb} used the measured visibilities to determine the size and geometry of the disk around \object{MWC~147} across K-band and N-band, combining their visibility measurements with a fit to the observed spectral energy distribution (SED).
They found that the best fit to the SED and the interferometry data was achieved using a model of a dust disk with an optically thick gaseous component inside the dust sublimation radius.
This model suggests that the K-band emission is dominated by the accretion luminosity from the active, gaseous inner disk, whilst the MIR emission still contained the flux contributions from the irradiated dust of the disk rim.\\
\citet{2010Bagnoli} obtained optical spectroscopy on \object{MWC~147} in several emission lines including forbidden and permitted OI and MgII transitions. The highly symmetric, double peaked [OI] line profiles suggest that the emission arises from a rotating circumstellar disk.
By deconvolving the line profiles and determining the radial intensity profile of the line-emitting gas they determined that the disk transitions from a gaseous disk to a dusty disk at radii of ${\sim}2-3$\,au, corresponding to the expected dust sublimation radius predicted by \citet{2008Krausb}.
Similar analysis of the Mg[II] emission lines suggested an inner radius of the gas disk at ${\sim}0.1$\,au, close to the expected co-rotation radius.\\

\citet{2014Ilee} observed CO bandhead emission around several Herbig Ae/Be stars, including \object{MWC~147}, using the XSHOOTER and CRIRES instruments at the VLT.
They found that the CO emission was located in a disk ranging from $0.89$ to $4.3$\,au with an inclination of $52^\circ$, in good agreement with the result from \citet{2008Krausb}.
A recent interferometric study by \citet{2017Lazareff} used the VLTI/PIONIER instrument in the H-band to analyse the circumstellar environment around several Herbig B[e] stars including \object{MWC~147}.
They were able to construct a detailed geometric model of the near-infrared emission, finding that it arose from a disk oriented with an inclination of $23\pm10^\circ$, i.e.\ more face-on than the earlier estimates.
\citeauthor{2017Lazareff} also estimated a disk major-axis PA of $44^\circ$ but, with an uncertainty of $42^\circ$, were unable to place strong constraints on this value, a problem that is possibly compounded by the low (close to face-on) disk inclination.\\
The large-scale structures around \object{MWC~147} were studied by \citet{2014Li} who utilized deep MIR imaging to observe the surrounding diffuse nebula.
In their study, \citeauthor{2014Li} found that the shape of the nebula is highly asymmetric, with a bow-like structure extending from the central star along a PA of ${\sim}120^\circ$.
\citeauthor{2014Li} discussed the possibility that these complex filamentary structures could trace the inner walls of the cavity carved out by a bipolar outflow.\\

Using a variety of techniques, we aim to resolve the innermost regions of the protoplanetary disk of \object{MWC~147}, determine the nature of the continuum emission and probe the origins of the Br$\gamma$ emission that arises in the inner disk.
Combining interferometry data from multiple instruments we achieve the highest angular resolution interferometric observations of the \object{MWC~147} disk so far, interpreting our data with a geometric disk model in order to measure the orientation of the \object{MWC~147} disk system.
We also combine medium and high resolution spectro-interferometry with CRIRES spectroscopy to build an overview of the origin and kinematics of the Br$\gamma$-emitting gas, using a model-independent photocentre shift analysis as well as our kinematic modelling code.



\section{Observations and data reduction}
\label{Sec:obs}

\begin{table*}
\small
\caption{Observation log for our \object{MWC~147} data taken with VLTI/AMBER and VLT/CRIRES.
The numbers in the calibrator column correspond to the following stars with their K-band uniform disk diameters in parentheses: (1) \object{HD\,53510} (2.049\,mas), (2) \object{HD\,43023} (0.940\,mas), (3) \object{HD\,47127} (0.411\,mas), (4) \object{HD\,50277} (0.356\,mas), (5) \object{HD\,47157} (0.276\,mas), (6) \object{HD\,52456} (0.271\,mas), (7) \object{HD\,47575} (0.225\,mas), (8) \object{HD\,45638} (0.262\,mas), (9) \object{HD\,48977} (0.119\,mas). The NDIT/Pointings column gives the NDIT for all observations except for CHARA/CLIMB where it gives the number of pointings taken on the night. The baseline and PA columns give the average value across all pointings.
}
\label{tab:obslog}
\centering
\vspace{0.1cm}
\begin{tabular}{c c c c c c c c c}
\hline\hline
Instrument & UTC Date & Telescopes & UT  & DIT & NDIT/Pointings & Proj. baselines & PA & Calibrator\\ 
& & & [h:m] & [ms] & \# &  [m] & [$^\circ$] & \\  \hline 

AMBER HR-K & 2010-12-18 & U2/U3/U4 & 06:15 & 1000 & 2100 & 43.5\,/\,59.2\,/\,89.2 & -133.5\,/\,-73.4\,/\,-98.4 & 1 \\ 
\hline
AMBER MR-K & 2008-12-14 & U1/U3/U4 & 07:29 & 500 & 1560 & 54.6\,/\,82.1\,/\,128.5 & -144.2\,/\,-103.3\,/\,-119.4 & 2\\
 & 2009-12-04 & U2/U3/U4 & 07:58 & 500 & 840 & 46.2\,/\,84.2\,/\,50.7 & -133.5\,/\,-102.3\,/\,-74.1 & 3\\
\hline
AMBER LR-K & 2008-12-15 & U1/U3/U4 & 07:38 & 26 & 6000 & 101.0\,/\,47.2\,/\,128.0 & -139.5\,/\,-73.9\,/\,-119.9 & 4\\
\hline
CHARA/CLIMB & 2010-12-01 & S2/W2/E2 & - & 21 & 3 & 162.5/218.8/154.6 & 20.6/-114.5/-246.6 & 5,6 \\
 & 2010-12-22 & S2/W1/E2 & - & 16 & 4 & 223.5/304.3/247.0 & -14.8/-227.3/-76.4 & 5 \\
 & 2012-11-28 & S2/W2/E2 & - & 20 & 3 & 218.3/304.9/120.2 & 2.8/-195.8/-51.1 & 7,8 \\
\hline
CRIRES & 2010-10-26 & U1 & 06:40 & 60000 & 36 & N/A & N/A & 9 \\
\hline \hline
\end{tabular}
\end{table*}

\begin{figure}
	\includegraphics[width=\columnwidth]{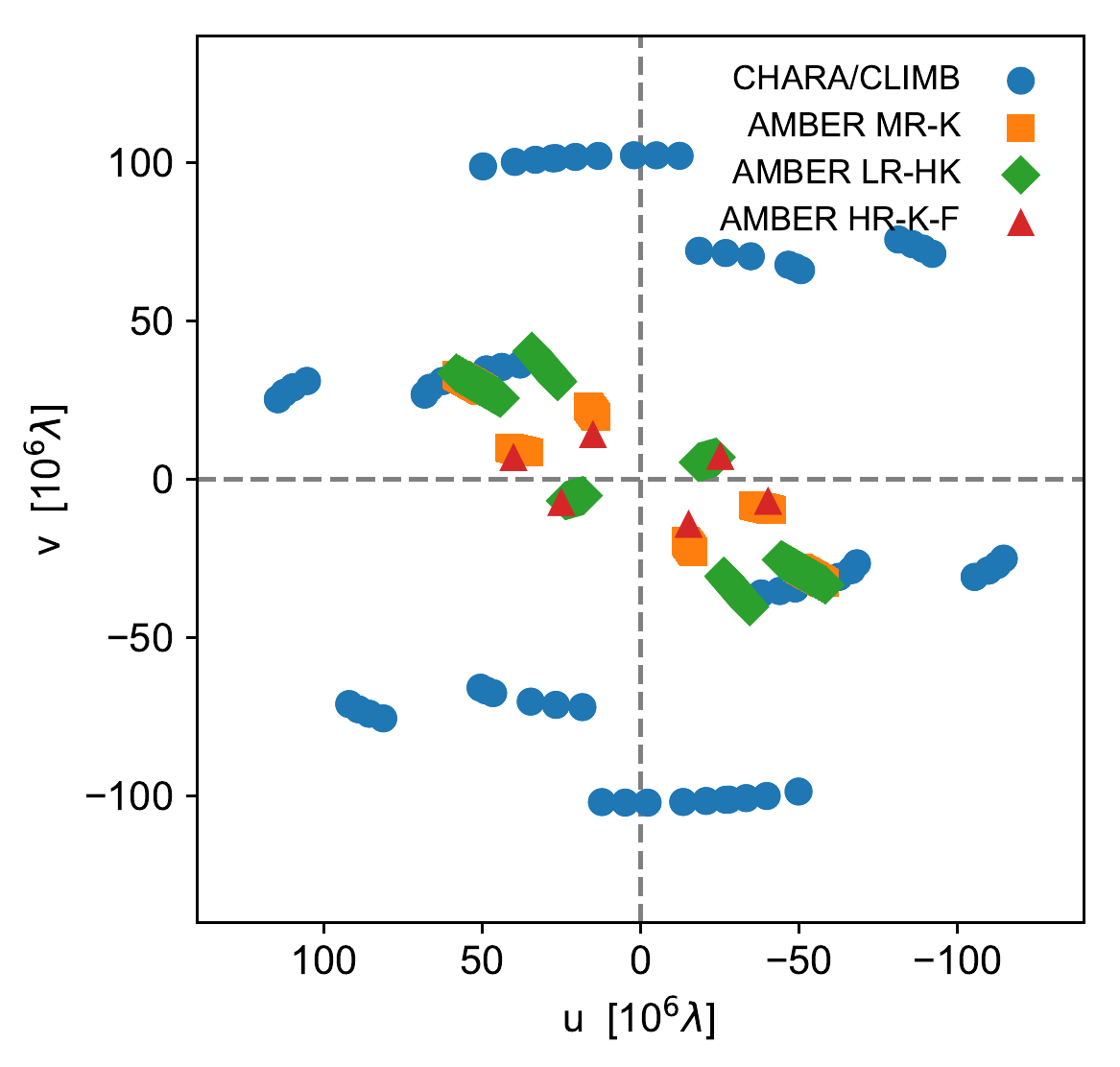}
    \caption{$uv$-coverage achieved with our interferometric observations of \object{MWC~147} with both VLTI/AMBER and CHARA/CLIMB.}
    \label{fig:uvcov}
\end{figure}

\subsection{VLTI/AMBER interferometry}

We observed \object{MWC~147} between 2008 and 2010 using the AMBER instrument at the VLTI.
AMBER \citep{2007Petrov} is a three telescope beam combiner capable of obtaining spectrally dispersed interferometric data with a range of spectral resolutions ($R=30$ to $R=12\,000$).
Our VLTI/AMBER observations are outlined in Table \ref{tab:obslog} and the $uv$-coverage of these observations is shown in Fig.\ \ref{fig:uvcov}.\\
Two of our data sets were taken with AMBER's HR ($R=12\,000$) mode in the K band, which allows us to spectrally and spatially resolve the Br$\gamma$ line from \object{MWC~147}.
Both of our AMBER HR-K datasets use the UT1-UT2-UT4 telescope triplet which covers a range of baseline lengths from ${\sim}56$\,m to ${\sim}130$\,m.
The first data set (2009-12-31, UT1-UT2-UT4) was recorded without external fringe tracking, but resulted only in a poor SNR.
Therefore, we reject this data set from further analysis.
For the second observation (2010-02-18, UT2-UT3-UT4) we used the FINITO fringe tracker \citep{2009LeBouquin} in order to correct for atmospheric turbulence, which enabled integration times up to 1 second and higher SNR.
In addition to the high spectral dispersion observations we also use data taken with AMBER's medium resolution ($R=1,500$, for which we also used FINITO) and low resolution ($R=30$) mode.
Data reduction was performed using our in-house AMBER data-processing software package, which uses the pixel-to-visibility matrix algorithm P2VM, as implemented in amdlib3 \citep{2007Tatulli, 2009Chelli} to extract wavelength dependent visibilities, differential phases, and closure phases.
We correct for heliocentric line-of-sight velocities using the ESO Airmass tool.
The visibility level of the continuum emission is calibrated using the best geometric model discussed in Section \ref{Sec:2Dmodel}, a necessary step due to the unreliability of visibilities that are observed with FINITO.
We detect a strong Br$\gamma$ line (${\sim}70\%$ above the continuum flux level) which causes the error bars in the line to be smaller than those in the continuum and as a result the continuum region has a poor SNR for our AMBER HR-K observations.

\subsection{CHARA/CLIMB interferometry}

The three-telescope beam combiner, CLIMB \citep{2013tenBrummelaar}, of the Center for High Angular Resolution Astronomy (CHARA) Array, was also used to obtain $K$-band continuum interferometry (see Table~\ref{tab:obslog}).
CHARA is an array of six $1\,$m-class telescopes with operational baselines between $34$ and $331\,$m \citep{2005tenBrummelaar}.
Observing \object{MWC~147} with the CHARA baselines gives us the opportunity to better resolve the continuum emission and achieve a better estimate of the inclination angle and position angle of the continuum disk, quantities that offer an important comparison with the perceived rotation angle of the Br$\gamma$-emitting gas.
\object{MWC~147} was observed on five separate occasions with CLIMB between 2010 December and 2012 November, achieving a maximum baseline length of ${\sim}300\,$m (corresponding to an angular resolution of ${\sim}0.75\,$mas).\\

The data were reduced using a pipeline developed at the University of Michigan that is well-suited to recovering faint fringes for low-visibility targets.
Standard stars not known to be members of binary or multiple systems were observed before and/or after each science observation and used to calibrate the visibilities and closure phases (see Table~\ref{tab:obslog} for the names and uniform disk [UD] diameters of the calibrators used\footnote{UD diameters were retrieved from JMMC SearchCal \citep{2006Bonneau, 2011Bonneau}}).
A further inspection of the closure phase signals of the calibrators observed more than once was undertaken to check for the presence of binary signatures and none were found.

\subsection{VLT/CRIRES spectroscopy}

In addition to our interferometry data we also have obtained high resolution ($R=100\,000$) spectral data for \object{MWC~147} in the K-band using the VLT/CRIRES instrument \citep{2004Kaeufl}.
Using CRIRES allows us to spectrally resolve the Br$\gamma$ line with the best possible spectral dispersion, allowing us to resolve the finer structure of the line which is crucial in order to distinguish between different kinematic line-emission scenarios.
Our observations of \object{MWC~147} were taken using a slit width of $0.2"$, 60-second integrations and using both the "nodding" and "jittering" techniques.
The nodding technique involves taking an integration on the object at an initial telescope position A, moving (or "nodding") the telescope to a second position B and taking two more integrations, then moving the telescope back to position A for a final integration.
"Jittering" involves adding a small random offset to each of the nodding offsets.
These techniques combine to remove sky emission, detector dark current and thermal noise as well as correcting for bad pixels and decreasing systematic errors.
The raw data was reduced using standard ESO pipeline in which the raw images are combined, spectra extracted and wavelength is corrected.
We correct for intrinsic Br$\gamma$ absorption in our calibrator (HD48977) before dividing the science spectrum by the calibrator spectrum.


\section{Analysis of interferometric data}
\label{Sec:AMBERLRMR}


\subsection{Geometric modelling of continuum emission}
\label{Sec:2Dmodel}

\begin{figure}
	\includegraphics[width=\columnwidth]{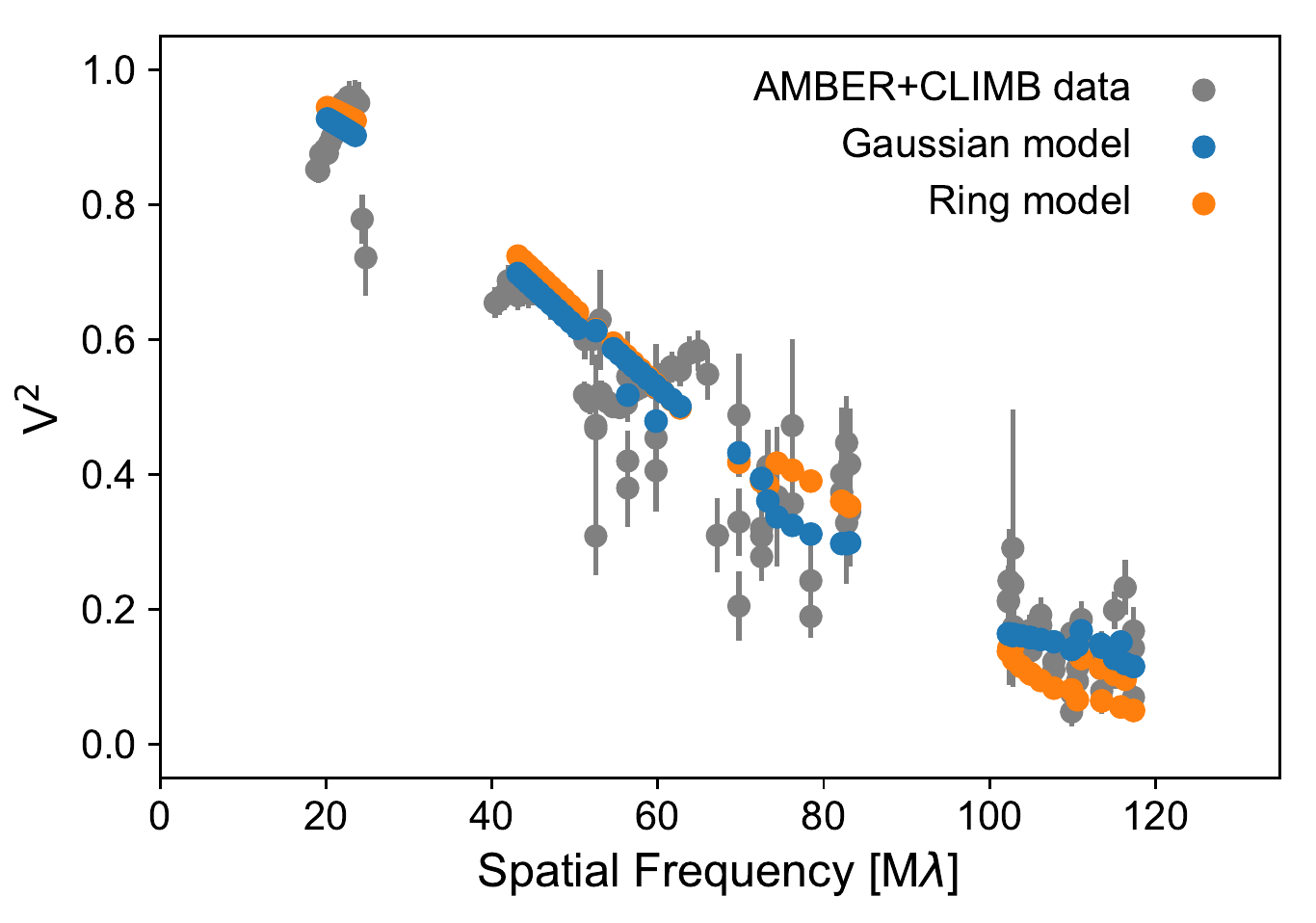}
    \caption{Visibility squared vs spatial frequency (baseline length) for our VLTI/AMBER and CHARA/CLIMB data for \object{MWC~147}.
    The observed data with uncertainties are shown in grey, whereas the results of our inclined Gaussian and ring models are shown in blue and orange respectively.}
    \label{fig:2dvis}
\end{figure}

\begin{figure}
	\includegraphics[width=\columnwidth]{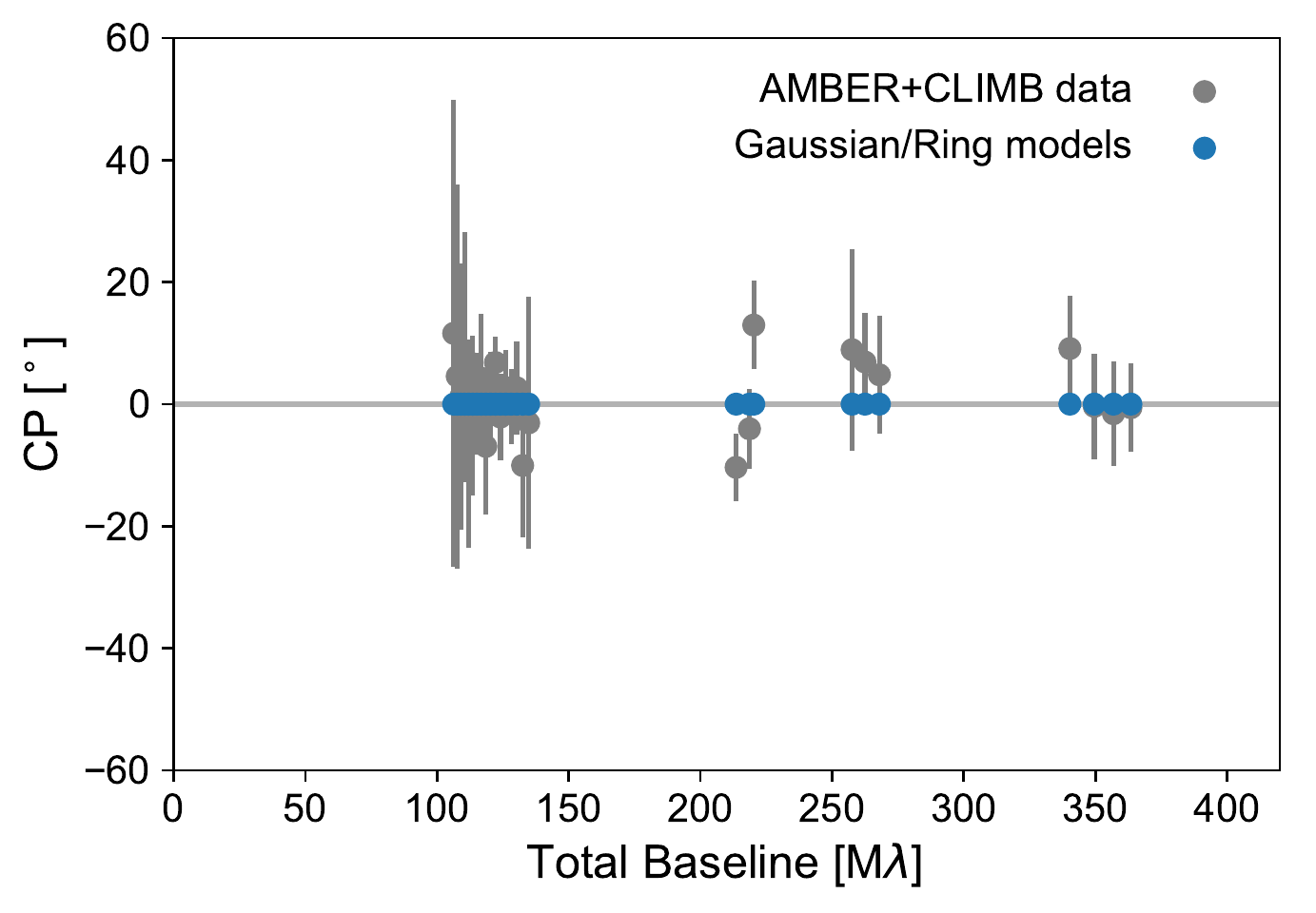}
    \caption{Closure phase vs combined baseline length for our VLTI/AMBER and CHARA/CLIMB data for \object{MWC~147}.
    The combined baselines length is calculated by summing the baselines for each closure phase triangle.
    The observed data with uncertainties are shown in grey, whereas the results of our inclined Gaussian and ring models are both shown in blue as they both have the same closure phase.}
    \label{fig:cp_model}
\end{figure}

\begin{figure}
	\includegraphics[width=\columnwidth]{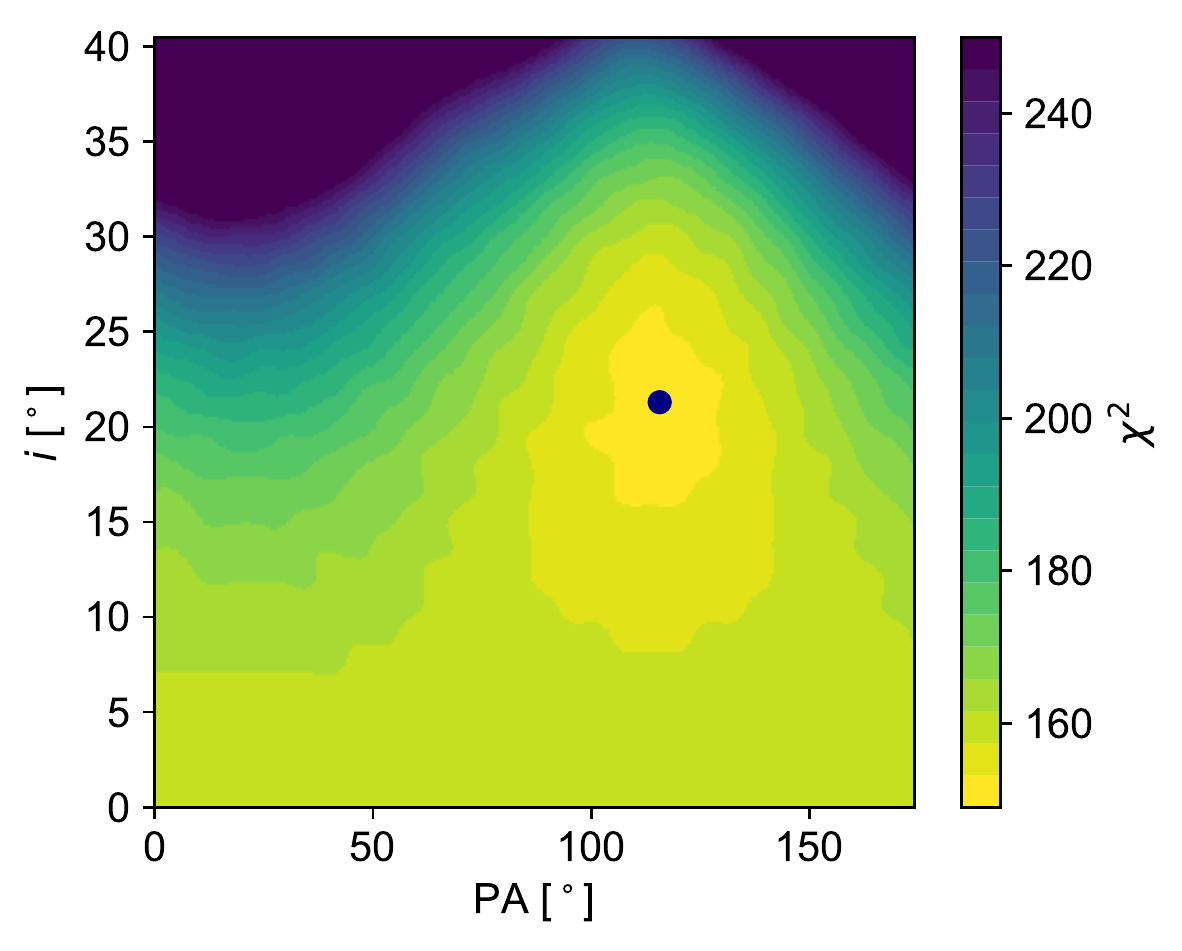}
    \caption{Two-dimensional $\chi^2$ map showing the dependence between the best-fit model PA and the inclination of our ring model. The color scale illustrates the total $\chi^2$ for each value of the PA and inclination and the best-fit model values are indicated by the blue data point.}
    \label{fig:2d_chi2}
\end{figure}

\begin{table}
\caption{Parameters for our best-fit geometric model featuring a star and inclined Gaussian or ring brightness distribution. $\theta$ indicates the PA of the major axis of the disk. The $\chi^2_r$ shown is calculated only using visibility data.}
\label{tab:2dparams}
\centering
\vspace{0.1cm}
\begin{tabular}{c c c c}
\hline \hline
Parameter & Range & Gaussian value & Ring value\\
\hline
$F_\star$ & - & $17\%{\pm6\%}$ & $16\%$\\
$F_{\text{disk}}$ & $0-20$ & $83\%{\pm6\%}$ & $84\%$\\
FWHM & $0$ - $5$ mas & $1.21{\pm 0.1}$ mas & -\\
R & $0$ - $5$ mas & - & $0.60\pm 0.03$ mas\\
dR & $0$ - $5$ mas & - & $0.2$R\\
$i$ & $0^\circ$ - $90^\circ$ & $21.3{\pm 6.4^\circ}$ & $26.7^\circ{\pm 4.7^\circ}$\\
$\theta$ & $0^\circ$ - $360^\circ$ & $25.7{\pm 15.7^\circ}$ & $42.4^\circ{\pm 15.0^\circ}$\\
\hline
$\chi^2_r$ & - & $1.53$ & $2.45$\\
\hline \hline
\end{tabular}
\end{table}

To determine the size and shape of the K-band continuum-emitting region we combine our CHARA/CLIMB and AMBER low spectral dispersion visibility data and interpret it with a simple geometric model.
Both CHARA and AMBER can also measure the closure phase, a quantity that is equivalent to the summation of the phase for each of the three baseline vectors in a closed triangle of baselines \citep{1958Jennison}.
The closure phase is a useful observable for detecting the presence of asymmetries in the measured brightness distribution \citep[e.g.][]{2003Monnier} and is an important component for obtaining higher-order information from interferometric observations.
However, we find that the measured data from AMBER and CLIMB is consistent with a null closure phase, with very few data points exhibiting a deviation from zero greater than 1$\sigma$ and zero deviations greater than 2$\sigma$ (see Fig.~\ref{fig:cp_model}).
As such we model the visibilities and closure phases from AMBER and CLIMB with centro-symmetric geometric models and achieve a $\chi^2_r$ of $0.5$ for our CP data.\\

Using the Rayleigh criterion, $\theta = \frac{\lambda}{2B}$, we determine that our maximum baseline allows us to achieve an angular resolution of $0.68$\,mas.
A two-dimensional elliptical Gaussian model is sufficient for our initial fitting where we focus on estimating inclination and PA, with the precise geometry not important for this goal as our baseline range only covers the first lobe of the visibility function (model parameters shown in Table~\ref{tab:2dparams}).
In addition we fit a simple ring model to our visibility data in order to compare the modelled ring radius with the expected dust sublimation radius.
The uncertainties of our best-fit parameters are determined via bootstrapping \citep[for a full description of this process see][]{2018Davies}.\\

In our Gaussian model we vary the full width at half maximum (FWHM), the disk major axis PA ($\theta$), inclination ($i$) and the relative flux between the disk and star.
We find that our best-fit model has a FWHM of $1.21\pm0.10$mas, which corresponds to $0.84{\pm0.074}$au assuming a distance of $711\,$pc.
We find that there is a loose dependence between the best-fit PA and inclination (illustrated by the $\chi^2$ map in Fig.~\ref{fig:2d_chi2}).
Figure~\ref{fig:2d_chi2} also shows a strong degeneracy in the PA value at low inclinations, possibly due to the low inclination that we observe for the disk.
Our best-fit model has a stellar flux contribution ($F_\star/F_\text{Tot}$) of $17\%{\pm}6\%$ which is consistent with the stellar flux contribution value of $16\%$ determined from the SED in \citet{2001Millan-Gabet,2008Krausb}.\\

We also modelled our VLTI+CHARA visibility data set with a two-dimensional ring brightness distribution.
This model is similarly constructed to our Gaussian model, with the same use of the inclination, PA and disk flux parameters, but with a ring of radius $R$ as free parameter and constant ring width of $0.2R$ \citep{2005Monnier}.
The fit yields a ring radius of $R=0.84$mas and an (unrealistically) high stellar flux contribution ($49\%$ of the total flux).
We also fit the visibility data with a ring model in which we allow the relative width of the ring to vary, which results in similar values for the best-fit PA and inclination as the model with fixed ring width (within $1\sigma$).
However, the algorithm pushes the fit towards a Gaussian-like brightness distribution, with a ring radius of $0.12$\,mas, a ring width of $9.9R$ ($\sim1.2$\,mas) and a stellar flux contribution of $\sim42\%$.
Both of the aforementioned models appear to strongly overestimate the stellar flux contribution compared to analyses that tried to separate the star/disk flux contributions from the SED \citep[$16\%$][]{2001Millan-Gabet,2008Krausb}.
Likely this is due to a well-known degeneracy that exists between the best-fit ring radius and stellar flux contribution when fitting visibility data in the first lobe of the visibility function \citep{2003Lachaume,2017Lazareff}. 
Therefore, we also conducted fits where we fix the stellar flux contribution to the SED value.
Assuming again a fixed-width ring of $0.2R$, we derive a ring radius of $0.60{\pm0.01}$\,mas (see Table~\ref{tab:2dparams} for full list of best fit parameters).
We adopt this model as our best fit ring model, although it is important to note that this model does not account for any other potential sources of flux other than the ring and therefore can only be considered an approximation to the true brightness distribution.\\

Our best-fit ring radius of $0.60{\pm0.01}$\,mas corresponds to a physical distance of $0.39{\pm0.01}$au assuming that the distance to \object{MWC~147} is $711\,$pc.
This region of the disk is $3.9$ times more compact than the expected \object{MWC~147} dust sublimation radius of $1.52$\,au assuming a dust sublimation temperature of 1800~K.
Our Gaussian model achieves a better fit to the visibilities than our ring model (see the $\chi^2_r$ values in Table~\ref{tab:2dparams}) although both models feature similar values for inclination and PA, meaning that each model finds a similar best-fit disk orientation.
The improved $\chi^2_r$ for our Gaussian model is reflected by the results of our variable-width ring fit which tends towards a Gaussian-like brightness distribution, suggesting that our Gaussian model is the most accurate representation of the true brightness distribution that we find in our modelling.
Our best estimate for the orientation of the MWC147 system is an inclination of $19.2\pm1.7^\circ$ and a major axis position angle of $12.7\pm12.1^\circ$, parameters that are derived from our kinematic model but are also consistent with our geometric modelling of the system.\\


\subsection{Photocentre analysis of Br$\gamma$ emission}
\label{Sec:photocentre}

\begin{figure*}
    \centering
	\includegraphics[width=130mm]{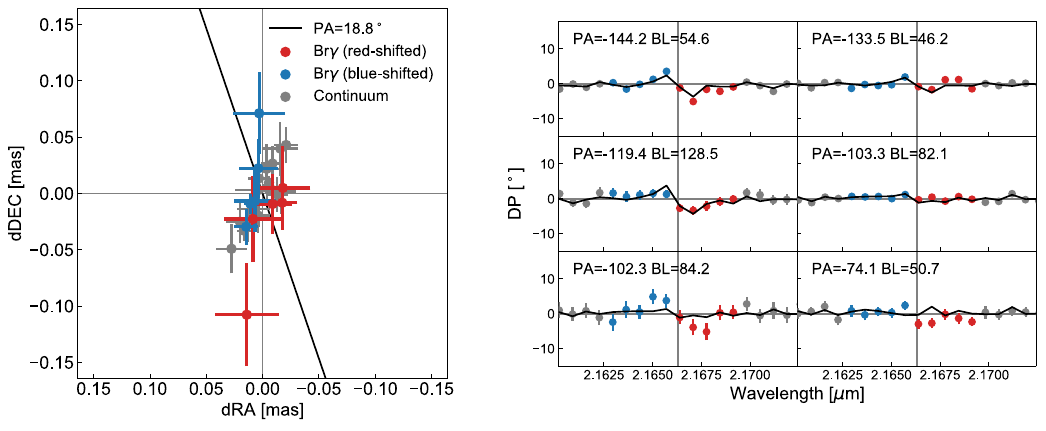}
    \caption{Left: Derived 2D photocentre displacement vectors for the Br$\gamma$ line + continuum determined from our AMBER MR-K data. The black line shows and estimated PA for the displacement between blue and red-shifted vectors.
    Right: Differential phase (DP) data for \object{MWC~147} observed with AMBER's MR-K mode (circular points) compared with differential phases calculated from the photocentre shifts shown in the right panel of this Figure (black lines). The baseline label in the upper corner gives the PA in degrees and the baseline (BL) in metres. The line (red and blue) and continuum (grey) points from the left panel correspond to the similar coloured points in the right panel.}
    \label{fig:photocentre}
\end{figure*}

We detect that for many of our observed baselines the differential phase deviates from zero in the Br$\gamma$ line (such as can be seen in Fig.~\ref{fig:simmap-model}), an effect that is caused by the centre of the emission in a particular wavelength channel deviating from the centre of the continuum emission.
By qualitatively analysing the differential phase patterns across the Br$\gamma$ line we can glean information about the differing kinematic scenarios traced by the line emission.
The "S-shaped" differential phases that we can see for several of our baselines (see Fig.~\ref{fig:photocentre}, right panel) are an indicator of rotation, as we see that the blue shifted emission is symmetrically displaced opposite to the red-shifted emission.
By combining the differential phases for multiple baselines and position angles we can derive a 2D photocentre displacement profile which reveals the position angle of the rotation traced by the line emission.
These photocentre displacements can be compared with the continuum disk geometry that we measure in the previous section and can help to constrain the PA of the disk major axis.\\

To derive the wavelength-dependent photocentre shifts from our differential phase data we use the fitting method outlined in \citet{2017Hone} that is based on the equation:

\begin{equation}
\label{eq:photocentre}
\vec{p} = -\frac{\phi_i}{2\pi}\cdot \frac{\lambda}{\vec{B_i}},
\end{equation}

\noindent where $\phi_i$ is the differential phase measured for the $i$th baseline, $B_i$ is the corresponding baseline vector and $\lambda$ is the central wavelength \citep{2003Lachaume, 2009LeBouquin}.\\

Examining our photocentre displacement vectors, we see a displacement between the blue and red-shifted vectors along a PA of ${\sim}18.8^\circ$ (see Fig.~\ref{fig:photocentre}).
This angle is estimated by determining the mean displacement for the blue shifted and red shifted vectors and calculating the displacement angle between the two mean vectors.
The extension of the data points in the north-south direction is caused by the non-uniform uv-coverage, and is oriented along the direction of least resolution (see Fig.~\ref{fig:uvcov}).
Our geometric model of the continuum disk indicates a disk major axis PA of $25.7{\pm 15.7^\circ}$ for the continuum-emitting disk which is consistent with the estimated rotation angle derived from our photocentres.
In order to construct a more detailed interpretation of both the differential phases and visibilities of the AMBER MR and HR data we use, in the next section, a kinematic model.


\section{Kinematic modelling}
\label{Sec:kinematicmodel}

\begin{figure*}
	\includegraphics[width=18.4cm]{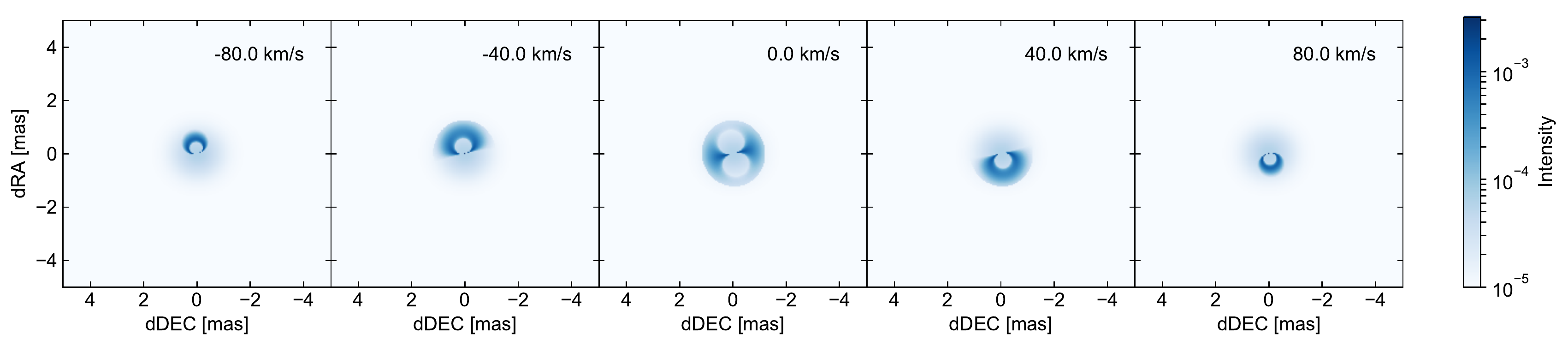}
    \caption{Synthetic model images calculated for our Keplerian accretion disk model for five spectral channels across the Br$\gamma$ line. We use an arbitrary logarithmic color scale to show the flux distribution with more detail. The parameters for this model can be found in Table~\ref{tab:modelparameters}.}
    \label{fig:modelsummary}
\end{figure*}

\begin{figure*}
	\includegraphics[width=18.4cm]{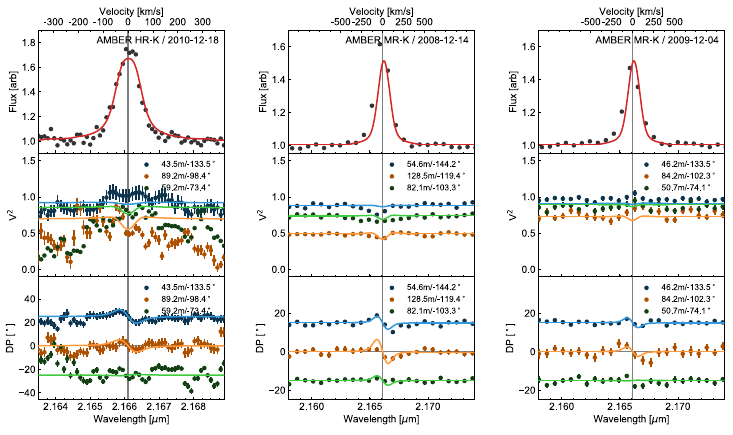}
    \caption{
    Comparisons of our observed interferometric data (coloured points) with synthesized data from our kinematic model (solid lines).
    Our AMBER HR data is shown on the left panel, with our two AMBER MR epochs in the central and right panels.
    Each plot shows the spectrum (top), squared visibilities (middle) differential phases (lower) calculated from our kinematic model and compared to our observed VLTI/AMBER data.
    The different coloured lines correspond to the different observed baselines which are labeled in the upper right corner of the panel.
    }
    \label{fig:simmap-model}
\end{figure*}

\begin{figure}
	\includegraphics[width=\columnwidth]{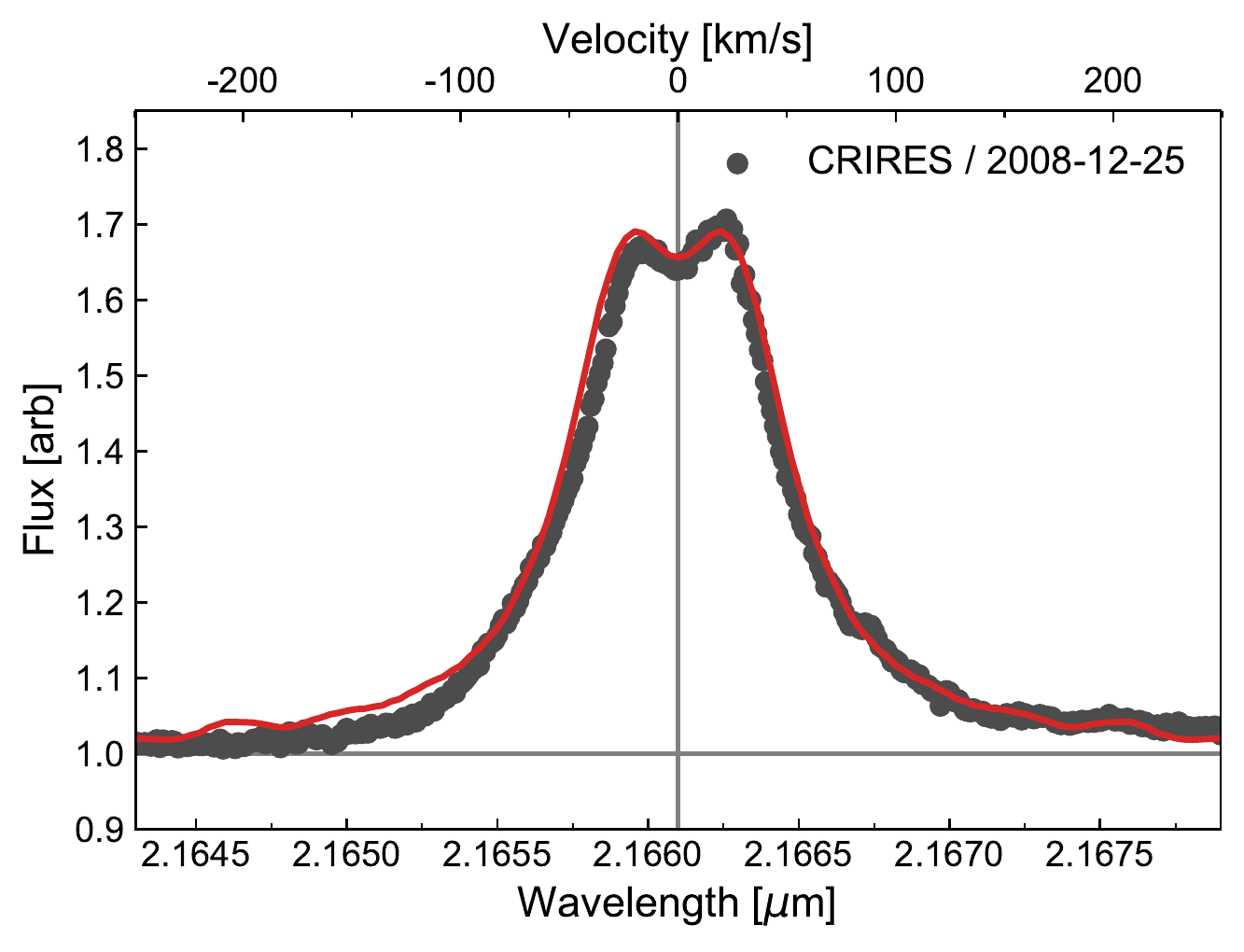}
    \caption{Spectrum calculated from our kinematic model and compared to our observed VLT/CRIRES data.
    The solid red line represents the model spectrum and the grey points represent the observed CRIRES spectrum.}
    \label{fig:crires-model}
\end{figure}


In order to constrain the gas velocity field around \object{MWC~147} we fit a kinematic model to our AMBER MR and HR data, as well as the CRIRES spectra.
We use the kinematic modelling code that we have already used for modelling spectro-interferometric observations for a range of evolved \citep{2007Weigelt,2012Krausb} and young objects \citep{2012Krausa,2017Hone}.
Based on an analytic description of 3D velocity fields and the radial brightness distribution, this code allows us to compute synthetic spectra and synthetic images for different velocity channels in a spectral line.
This type of modelling is a useful tool to explore how different disk morphologies and variations of the velocity field fit the observed visibilities and differential phases.\\

Using our kinematic modelling code we compute synthetic intensity distribution frames for each spectral channel that include contributions from both a continuum-emitting disk and Br$\gamma$-emitting component.
The continuum emission is modelled as an inclined two-dimensional Gaussian identical to the one determined in Sect.~\ref{Sec:2Dmodel}.
The line-emitting region extends from an inner radius, $R_{in}$, to an outer radius, $R_{out}$, and has a radial brightness following the power-law ${\propto}r^\beta$, where $r$ is the radius and $\beta$ is a power-law exponent.
The disk major-axis PA ($\theta$) and inclination ($i$) are treated as free parameters with starting values taken from our fitting of the continuum disk.
To avoid unphysical sharp edges in our model images we use a Fermi-type smoothing function to model the edges of the disk \citep[see][]{2008Krausb}.
We adopt the stellar parameters (mass $M_{\star}$ and distance $d$) quoted in Sect.~\ref{Sec:intro}.
From the resulting 3D image cube we compute the wavelength-dependent interferometric quantities for the $uv$-coordinates covered by our data.\\

In order to determine which kinematic scenarios to explore we evaluated our observed AMBER and CRIRES data.
We do not see a significant variation in visibility across the line, indicating that the Br$\gamma$ emission originates in a similar compact region as the continuum disk.
The innermost stream lines of the disk wind region (and the X-wind region) may be located near the corotation radius (${\sim}0.1$au), therefore the measured radius of ${\sim}1$au of the Br$\gamma$-emitting region is consistent with the expected emission region of a disk wind.
Additionally, the observed CRIRES spectrum (shown in Fig.~\ref{fig:crires-model}) shows a highly symmetric, double-peaked Br$\gamma$-line profile which is indicative that the line emission arises in a velocity field that is either purely rotational or does not contain a significant out-of-plane velocity component.
This suggests that the emission arises from this disk surface or a region very close it, where material ejected by a possible disk wind is not yet significantly accelerated and therefore the velocity field is still approximately Keplerian.
Taking our conclusions from analysis of this data, we elect to model the Br$\gamma$ emission around \object{MWC~147} with the  simpler case of a gaseous disk in Keplerian rotation.\\

Our model is at first calculated with a spectral resolution comparable to that of CRIRES (as it is our highest spectral resolution data), so to compare our model quantities to our lower spectral resolution AMBER data we convolve our model to spectral resolutions of AMBER's HR and MR modes with a simple Gaussian kernel.
Taking the results from Table~\ref{tab:2dparams} as initial values for our kinematic model parameters we construct a series of model grids, taking the model with the lowest $\chi^2_r$ as our best-fit model, the parameter values of which are shown in Table \ref{tab:modelparameters}.
The best-fit model brightness distribution for five velocity channels can be seen in Fig.~\ref{fig:modelsummary} and the comparison between our observations and the synthesized observables from the model can be seen in Figures \ref{fig:simmap-model} and \ref{fig:crires-model}.
After modelling the continuum disk with the same parameters from our AMBER/CLIMB fitting in Sec.~\ref{Sec:2Dmodel}, we tweak the parameters of the line-emitting region in order to find the best possible fit to the observed CRIRES spectrum and AMBER spectrum/visibilities.
The precise shape of the model spectrum is highly sensitive to the disk inclination as well as the radial brightness distribution of the line-emitting region.
By adjusting these parameters of our kinematic model we can match the full width of the high velocity spectral line wings as well as the double-peak structure that we see at very low velocities (in the line centre, see Fig.~\ref{fig:crires-model}).
We alter the disk position angle in order to best recreate the patterns we see in the differential phase data.
The uncertainties of our model variables are calculated using the $\chi^2$ method.
Our kinematic model can reproduce the observed CRIRES spectrum with a reduced $\chi^2_r$ of $1.44$ and the parameters of the model are listed in Table~\ref{tab:modelparameters}.\\

The differential phase measurements from our AMBER MR and HR data are reproduced with moderate success by our kinematic model.
In particular, the overall shape of the model differential phase patterns is a good match to the observed data, but the magnitude of the phase is marginally too high or low for some baselines.
The visibility levels for the two MR data sets are well reproduced (Fig.~\ref{fig:simmap-model}, middle and right panels), although the changes in visibility across the line (whilst very small) are not well matched for our 2009-12-14 data.
This could be due to inconsistencies between the data sets caused by intrinsic variability of the source, which we do not account for in our models.
This is also reflected by the different line strengths shown in the two MR data sets.
Our model achieves a $\chi^2_r$ of $0.56$ for the visibilities ($\chi^2_{r,V}$) and a $\chi^2_r$ of $0.33$ for the differential phases ($\chi^2_{r,\phi}$) leading to a total $\chi^2_r$ of 0.89.
Whilst our model is a good match to the differential phases of our AMBER HR data set (Fig.~\ref{fig:simmap-model}, left panel), the visibilities for this data set have a very low SNR and there are few coherent patterns visible for our model to replicate.
The strength, width and fine structure of the observed CRIRES ($R=100\,000$) spectrum is very well reproduced by our kinematic model, but the slight variation that we see in the line strength for our different AMBER MR and HR data sets means that the fit to the AMBER spectra is good but not perfect.


\begin{table}
\caption{Ranges and best-fit values for our kinematic model parameters. The reduced $\chi^2$ values (calculated using our AMBER HR data) for each of our best-fit kinematic models are also shown.}
\label{tab:modelparameters}
\centering
\vspace{0.1cm}
\begin{tabular}{c c c}
\hline \hline
Parameter & Range & Best-fit value\\
\hline
$R_{in}$ & $0$ - $5$ mas & $0.1{\pm 0.04}$ mas\\
$R_{out}$ & $0$ - $10$ mas & $2.1{\pm 0.5}$ mas\\
$\theta$ & $-180^\circ$ - $180^\circ$ & $12.7{\pm 12.1^\circ}$\\
$i$ & $0^\circ$ - $90^\circ$ & $19.2{\pm 1.7^\circ}$\\
$\beta$ & $-2.0$ - $0.0$ & $-0.8{\pm 0.1}$\\
\hline
$\chi^2_{r,V}$ & - & $0.56$\\
$\chi^2_{r,\phi}$ & - & $0.33$\\
$\chi^2_{r}$ & - & $0.89$\\
\hline \hline
\end{tabular}
\end{table}


\section{Discussion}


Based on the measured bolometric luminosity of $1549\,L_{\odot}$ \citep{2004Hernandez}, we expect that the dust sublimation radius should be located at $\sim1.52$\,au (assuming a sublimation temperature of $1800$K and grey dust opacities), corresponding to angular diameter of ${\sim}4$\,mas at the distance of $711$\,pc.
Our findings show that the K band continuum is much more compact, which suggests that the emission does not originate from the dust sublimation rim, but an alternate source located about 3.4 times closer in.
This occurrence of "undersized" Herbig Be stars has been observed in several other high-luminosity objects in the size-luminosity study of \citet{2002Monnier}.
The compact nature of these objects has been interpreted in several different ways, including an inner gaseous component that shields stellar radiation to let dust survive closer to the central star \citep{2002Monnier,2004Eisner,2011Weigelt} or additional emission components such as highly refractory dust grains \citep{2010Benisty}.\\
In the particular case of \object{MWC~147}, \citet{2008Krausb} suggested that the K-band continuum traces optically thick gas from a compact, viscously-heated accretion disk.
\citeauthor{2008Krausb} combined NIR and MIR interferometric data for \object{MWC~147} and found that models with a passive dust sublimation front were not able to reproduce the measured steep increase in the characteristic size of the continuum-emitting region (factor $\sim10$ between 1 and 12\,$\mu$m).
However, they were able to obtain a good fit to the visibilities, closure phase and SED by including an optically thick, gas accretion disk inside of the dust sublimation rim.
The presence of hot accreting gas in the inner disk is also consistent with \object{MWC~147}'s high accretion rate \citep[{}$\dot{M}_{\text{acc}} \approx10^{-5}\,M_\odot$yr$^{-1}$;][]{1992Hillenbrand}.
Measurements of the differential visibility do not show a consistent, significant variation across the Br$\gamma$ line, suggesting that the line emission originates from a similar region to the K-band continuum.
The presence of line-emitting gas that is co-located with the continuum emission is strong evidence that the continuum traces an active, gaseous accretion disk inside of the dust sublimation radius \citep[as was suggested by ][]{2008Krausb, 2010Bagnoli}, rather than an irradiated disk wall or highly refractory dust grains.\\

Combining our geometric and kinematic models we attempt to obtain a precise estimate for the disk PA, a quantity which has been poorly constrained by previous studies using geometric models \citep{2008Krausb,2017Lazareff}, possibly due to the close to face-on inclination of the disk.
We constructed model-independent photocentre shifts using our AMBER MR differential phase data in order to determine the rotation angle of the Br$\gamma$-emitting gas.
Our photocentre vectors show displacement from the central continuum region along an estimated PA of ${\sim}18.8^\circ$, in a similar direction to the major axis PA of our best-fit geometric model ($25.7{\pm 15.7^\circ}$).
Using our kinematic model, we derive a disk inclination of $19.2^\circ{\pm1.7}$ for the inner gas disk around \object{MWC~147}, which is consistent with our estimate for the inclination of the continuum-emitting disk ($21.3{\pm5.8}^\circ$), derived from our geometric modelling of the continuum-emitting disk using AMBER/CLIMB visibility data.
Our kinematic model of the gas disk has a major axis PA of $12.7^\circ{\pm12.1}$.
Our best-fit major axis PA for the continuum-emitting disk is $25.7^\circ\pm15.7$ which is consistent with the value determined by our best-fit kinematic model of the gas disk, indicating that the Br$\gamma$ emission traces Keplerian rotation in a gas disk which lies in the same plane as the continuum-emitting disk.
Our best disk PA (determined from our kinematic model of the gas disk) of $12.7^\circ{\pm12.1}$ gives the most robust estimate of the disk PA yet. Using AMBER differential phases and the long CHARA baselines we can significantly improve on the constraints from \citet{2017Lazareff}, who estimated the major axis PA to be between $2^\circ$ and $86^\circ$.\\

Using CRIRES spectroscopy we spectrally resolve the Br$\gamma$ line from \object{MWC~147} with very high resolution, examining the fine structure and comparing the line profile with similar objects to learn more about the origin of the line emission.
The shape and width of the Br$\gamma$ line that we observe is very similar to the shape and width of the [OI] optical transitions observed by \citet{2010Bagnoli}, suggesting that the two different species are present in a similar region of the disk.
Our CRIRES data ($R = 100\,000$) reveals that the line has a highly symmetric double-peaked structure, with line wings that extend out to velocities of $\pm200$\,km/s.
This provides an interesting comparison point to our previous study of \object{MWC~297} \citep{2017Hone}, which has a similar disk geometry (${i=\sim}30^\circ$) albeit around a more massive object.
Typically, emission lines that arise from rotating disks exhibit double-peaked lines at moderate-to-high inclinations ($i{\gtrsim}20^\circ$) with the line becoming single-peaked at inclinations below ${\sim}20$.\\

In our \citeyear{2017Hone} paper, the CRIRES spectroscopy showed that \object{MWC~297} has a slightly asymmetric single-peaked Br$\gamma$ line despite having an inclination of ${\sim}23^\circ$, similar to the $24.8^\circ$ inclination of \object{MWC~147}.
\object{MWC~297} has a wider Br$\gamma$ line, which would be expected for a more inclined object, and the line is single peaked, suggesting emission process of the Br$\gamma$ line is different for \object{MWC~297} and \object{MWC~147} and that simple inclination effects cannot account for the differences in the line emission for the two objects.
The asymmetric line profile of \object{MWC~297} was well reproduced by a more complex disk-wind model whereas, for \object{MWC~147}, the strongly symmetrical observed Br$\gamma$-line suggests that the emission arises in a rotating circumstellar disk or a region of a disk wind near the disk surface where material has not been significantly accelerated.
The differences between these two scenarios are subtle \citep[as we discuss in][]{2017Hone} and it is likely only possible to distinguish between the two cases if there is a more significant out-of-plane velocity component present in the velocity field traced by the Br$\gamma$ emission.
Such out-of-plane velocities are found in disk wind scenarios in the outer regions where material is more accelerated but also more diffuse, leading to a much smaller contribution to the overall line emission.
We find that our model of a Keplerian disk is able to accurately reproduce the observed CRIRES Br$\gamma$ spectrum (with a reduced $\chi^2$ of $1.44$). 
It is important to note that with the AMBER HR spectral resolution ($R=12\,000$) the differences between these two line shapes would be indistinguishable from each other, highlighting the importance of using high resolution spectroscopy in conjunction with spectro-interferometry when modelling the line emission from YSOs.\\

Our interpretation of the Br$\gamma$ emission originating from a gaseous accretion disk is only a good explanation if it can be demonstrated that accretion is able to heat the gas in the disk to temperatures of $\sim6000-10\,000$K, hot enough to emit in the Br$\gamma$ line.
This question was addressed by \citet{2016Tambovtseva}, who investigated whether an optically thick gaseous accretion disk could contribute to the Br$\gamma$ emission in addition to a disk wind and magnetosphere.
Six disk models were computed with varying size, temperature and accretion rate, with the outcome of this modelling demonstrating the viability of the gaseous accretion disk scenario that we employ for out study of \object{MWC~147}.

\section{Conclusions}

In this paper we observed the inner regions of the protoplanetary disk around \object{MWC~147} with high spectral and spatial resolution.
Using VLTI/AMBER and CHARA/CLIMB interferometry, we can build up a picture of the inner disk ranging from the morphology of the continuum-emitting disk to the kinematics traced by the gaseous line-emitting disk.
We complement our interferometric data with very high resolution VLT/CRIRES spectroscopy of the Br$\gamma$ line.
We construct a multi-component kinematic model of an active gaseous accretion disk that provides a good fit to our AMBER and CRIRES data.\\

Using VLTI and CHARA interferometry, we probe the K-band continuum of \object{MWC~147} with $2$-times higher angular resolution than ever before, detecting a very compact continuum-emitting disk, corroborating the findings of previous works by \citep{2008Krausb,2017Lazareff}
Our ring model for the continuum emission has a radius of 0.60~mas ($0.39$~au) which is ${\sim}3.9$ times more compact than the expected dust sublimation radius of $1.52$~au.
By observing the rotation traced by the Br$\gamma$ emission from the gas disk and comparing the rotation axis with the observed position angle of the continuum-emitting disk, we are able to place the strongest constraints yet on the orientation of the \object{MWC~147} system.
Additionally, comparing the perceived Br$\gamma$ (gas-disk) rotation angle with the continuum-disk PA suggests that the two disks lie in the same plane, consistent with the model of the disk in Keplerian rotation rather than a more complex kinematic scenario such as a disk-wind.\\

We are able to achieve a very good fit to the double-peaked CRIRES Br$\gamma$ spectrum as well as the differential interferometric observables from our AMBER data with a kinematic model of a gaseous disk in Keplerian rotation. We find that both the gas emission and continuum emission arise from a compact region inside the expected dust sublimation radius.
The presence of this gas in the same location as the continuum emission strongly suggests that the compact K-band continuum emission arises from an active, viscous, gaseous accretion disk.
These findings provide direct evidence for compact gaseous accretion disks as an explanation for the phenomena of "undersized" Herbig Be stars.
In the future, other undersized Herbig Be stars should be studied with spectro-interferometry to determine whether line-emitting gas is co-located with the compact continuum emission similarly to \object{MWC~147}.

In order to more accurately measure the complex shape and and morphology of the \object{MWC~147} K-band continuum emission in the future, optical interferometry with baselines longer than ${\sim330}$m must be achieved in order to sample the second lobe of the visibility function.
Additionally, with the introduction of the GRAVITY instrument at VLTI, a 4-telescope spectro-interferometric beam combiner, the line emission around \object{MWC~147} can be resolved with twice as many baselines simultaneously and possibly a $uv$-coverage sufficient for spectrally-resolved image reconstruction can be achieved.

\begin{acknowledgements}
We acknowledge support from an ERC Starting Grant (Grant Agreement No.\ 639889), STFC Rutherford Fellowship (ST/J004030/1), Rutherford Grant (ST/K003445/1), and Philip Leverhulme Prize (PLP-2013-110).
This work is based upon observations obtained with the Georgia State University Center for
High Angular Resolution Astronomy Array at Mount Wilson Observatory.
The CHARA Array is supported by the National Science Foundation under Grant No.\ AST-1211929.
Institutional support has been provided from the GSU College of Arts and Sciences and the GSU Oce of the Vice President for Research and Economic Development.
This research has made use of the Jean-Marie Mariotti Center SearchCal service\footnote{Available at http://www.jmmc.fr/searchcal}, co-developed by LAGRANGE and IPAG.
\end{acknowledgements}

\bibliographystyle{aa}
\bibliography{MWC147}

\end{document}